\documentclass[epsfig]{aa}
\input psfig.tex
\usepackage{epsfig}
\begin{document}

   \thesaurus{03(11.19.1;11.19.3;13.09.1;13.25.2;11.09.1:IRAS~05189-2524)}
   \title{An X--ray and Near-IR spectroscopic analysis of the ULIRG
   IRAS~05189-2524 \thanks{Based on observations obtained at ESO, La Silla}}


   \author{P. Severgnini
          \inst{1}
	\and
	  G. Risaliti,  \inst{1}
	  A. Marconi,  \inst{2}
	  R. Maiolino \inst{2}
	    \and
	  M. Salvati   \inst{2}
          }

   \offprints{P. Severgnini, paolas@arcetri.astro.it}

   \institute{$^1$ Dipartimento di Astronomia, Universit\`a di Firenze,
    L.go E. Fermi 5, I--50125, Firenze, Italy\\
              $^2$ Osservatorio Astrofisico di Arcetri,
    L.go E. Fermi 5, I--50125, Firenze, Italy \\
 		}            

   \date{Received ; accepted}

 \titlerunning{X--ray and NIR analysis of IRAS~05189-2524}
 \authorrunning{P. Severgnini et al.}
 \maketitle

\maketitle

\begin{abstract}
We present new, quasi--simultaneous hard X--ray and near-IR spectra
of the ultraluminous infrared galaxy IRAS 05189-2524, and discuss them
together with archival and literature data.
The 1.9 Seyfert nucleus is Compton--thin. The near-IR broad lines
are seen in transmission, similarly to the X~rays, and the medium 
along the line of sight has an $A_V/N_H$ ratio definitely lower than 
Galactic. The increase in obscuration observed at the latter epoch
has $\Delta A_V/ \Delta N_H$ also less than Galactic, supporting
a correlation between the properties of the obscuring matter and its
proximity to the center. The measured
$A_V$ is compatible with the broad component of $H\alpha$ being seen
in transmission, as opposed to scattering, so that most of the observed
polarization must be due to dichroism.
The bolometric luminosity of the AGN, deduced from the X--ray and line
luminosities,
falls definitely short of accounting for the IR luminosity of the
galaxy, consistent with its coolish infrared color.

\end{abstract}

%
\section{Introduction}
The Ultraluminous Infrared Galaxies (ULIRGs) are sources of
quasar-like luminosities with a $\rm L_{IR} > 10^{12} L_{\odot}$
(H$_0$=75 km s$^{-1}$ Mpc$^{-1}$) much
higher than the infrared luminosity of normal galaxies 
($\rm L_{IR} \sim 10^{10} L_{\odot}$).
The high infrared emission is due to
the presence of large dust amounts absorbing, thermalizing and reemitting
the optical and UV photons of the primary source into the infrared band.
However, the nature of the primary optical-UV emission is still debated.
Starburst activity and/or an Active Galactic Nucleus (AGN) are the 
two mechanisms invoked as primary energy source.
Unfortunately most of the features commonly 
used to distinguish the two kinds of sources are often erased by dust,
and it is difficult to quantify the relative contribution of
starbursts and AGNs to the ULIRG phenomenon.
The studies conducted so far show that those ULIRGs which do not have
a Seyfert~1 or Seyfert~2 optical spectrum seem to be powered by
starbursts (Genzel et al. \cite{gen}, Rigopuoulou et al. \cite{rig},
Veilleux et al. \cite{vel}, \cite{vel1}).  
Moreover, even those ULIRGs hosting an AGN are in many cases
dominated by starburst emission. However, in
at least some of them the AGN could be  powerful enough to 
contribute significantly to the energy budget (Soifer et al. \cite{soi}, Vignati et al. \cite{vig}, Franceschini et al. \cite{fra}).
Nearly all ULIRGs appear to be advanced merger systems
(Goldader et al. \cite{gol}, Borne et al.\cite{bor},
Rigopuoulou et al. \cite{rig}), a fact consistent with both scenarios.\\
A recent comparison between the infrared and X--ray emission for all the
Luminous Infrared Galaxies (LIGs, L$_{IR}$$>$ 10$^{11}$L$_{\odot}$) observed in
the 2-10 keV band (Risaliti et al. \cite{ris}) suggests that these sources 
have reduced dust reddening and absorption with respect to what is expected from the gaseous N$_H$, assuming a Galactic dust-to-gas ratio and extinction curve.
The same result is also derived by comparing the optical and
X--ray emission of a sample of grism-selected QSOs (Risaliti et al.
\cite{ris1}), and of Seyfert galaxies of intermediate type (Maiolino et al.
\cite{mai1}). \\
IRAS~05189-2524 is a z=0.042 ULIRG selected from the 
Bright Galaxy Survey (BGS) of Sanders et al. \cite{san}
(L$_{IR}$=12.9$\times$10$^{11}$L$_{\odot}$,
Sanders \& Mirabel \cite{san:mir}, Risaliti et al. \cite{ris}). 
This source has been classified as a Seyfert~2 galaxy
on the basis of optical spectroscopy
(Veilleux et al. \cite{vel0}, Young et al. \cite{you}). At variance with 
the optical, the near-IR spectrum is rich of permitted broad
emission lines (Veilleux et al. \cite{vel1}).
Young et al. (\cite{you}) observed also a broad component of H$\alpha$,
but ascribed it to reflection and maintained the Sy~2 classification.
The same authors, fitting polarimetric and spectropolarimetric
observations, concluded that dichroic transmission is required in the
NIR, and deduced a relatively modest amount of absorption.
Instead, Clavel et al. (\cite{cla}) suggested that also the IR broad lines
could be due to reflection, because of the high degree of polarization 
exhibited by this source.

\noindent In the X rays, IRAS 05189-2524 is a relatively bright Compton 
thin source (Nakagawa et al. \cite{nak}, Risaliti et al. \cite{ris}):  
it is one of the brightest optically absorbed Seyfert 
galaxies observed in the hard X--ray band so far, and
the high X--ray flux together with the rich optical and IR emission line
spectra allow a detailed study of its nature.\\
In this paper we present an analysis of archival ASCA and new 
Beppo-SAX observations of IRAS~05189-2524 carried out in 1995 and 1999,
respectively.
The X--ray data are complemented with Near-InfraRed (NIR) 
spectra obtained with SOFI (NTT) in 1999, and with published 
UKIRT spectra obtained midway between ASCA and Beppo-SAX.
The observations, reduction and analysis in the X--ray  
and near-infrared regions are presented in Section~2 and Section~3, 
respectively. The main results we have obtained are discussed in 
Section~4 and the conclusions are reported in Section~5.
Throughout this paper we assume H$_0$= 75 km s$^{-1}$ Mpc$^{-1}$.


\section{X--ray Data} IRAS~05189-2524 
was observed by the X--ray observatory ASCA (Tanaka, Inoue \&
Holt \cite{tan}) from 1995 February 15 to 1995 February 16.
The source was observed again in 1999 October 3 with Beppo-SAX (Boella
et al. \cite{boe1}).

\begin{table}
\caption{ASCA and Beppo-SAX observations of IRAS~05189-2524.}
\label{obs_log}
\begin{tabular}{lcc}
\hline \hline   
& Exposure time& Net count rate\\
&  (s) & (cts/s)\\
\hline 
\multicolumn{3}{l}{\it ASCA(1995)}\\ 
\hline
SIS0& 77090& 0.043$\pm0.001$\\
SIS1& 77090& 0.035$\pm0.001$\\  
GIS2& 78230& 0.048$\pm0.001$\\
GIS3& 78230& 0.065$\pm0.002$\\
\hline 
\multicolumn{3}{l}{\it Beppo-SAX(1999)}\\
\hline  
LECS& 17415& 0.008$\pm$0.001\\
MECS& 42139& 0.037$\pm$0.001\\
\hline
\end{tabular}
\end{table}

\subsection{Timing analysis}
We have extracted the light curves of IRAS~05189-2524
from the ASCA and Beppo-SAX data using version 1.3 of XSELECT.
In order to increase the signal to noise ratio, we have combined the
SIS0 and SIS1 data and the GIS2 and GIS3 data.
The SIS detectors have operated in Bright 2 mode.
The light curves accumulated in the 2-10 keV band for the SIS and GIS data
are shown in the upper and lower panels of Fig.~1, respectively.
Fig.~2 shows the light curves accumulated 
from the LECS detector between 0.1 and 2 keV (upper panel) and from the MECS 
between 2 and 10 keV (lower panel). 
\begin{figure}
\epsfig{file=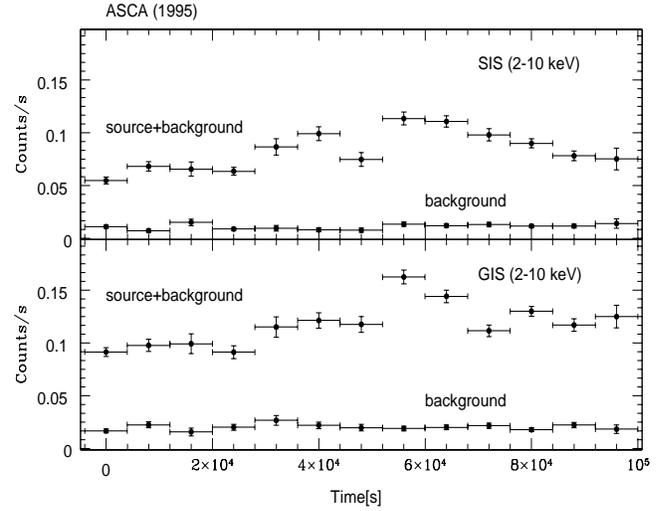, width=9.cm, height=10cm, angle=0}
\vskip -3truecm
\caption{Light curves for SIS (top) and GIS (bottom).
The binning time is 8~ks. The lights curves are not background
subtracted. The light curves of the background are also shown for comparison.}
\label{ascatiming}
\end{figure}
\begin{figure}
\epsfig{file=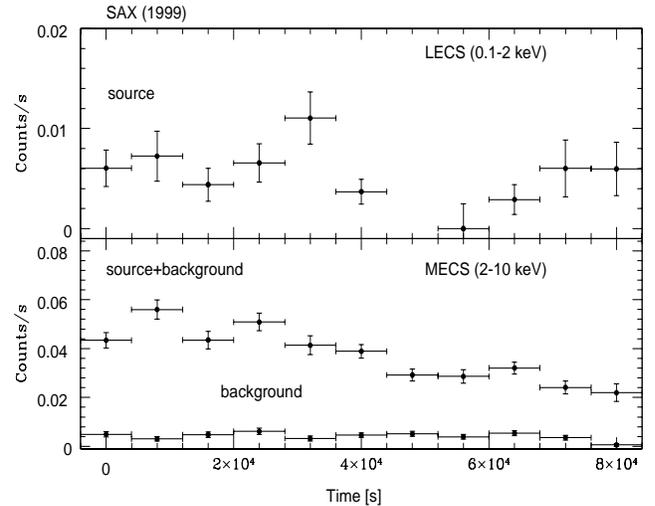, width=9.cm, height=10cm, angle=0}
\vskip -3truecm
\caption{Light curves for LECS (top) and MECS (bottom).
The binning time is 8~ks. The LECS curve is background subtracted
contrary to the MECS one.The light curve for the background in the 2-10 keV
band is also shown.}
\label{saxtiming}
\end{figure}
In the 2-10 keV band IRAS~05189-2524 exhibits short term flux 
variability both in the 1995 and 1999 observations.
In  1995 the flux varied by a factor of $\sim$ 2 on a timescale of
6$\times$10$^4$~s.
In 1999 the source exhibits the same level of variability, 
perhaps on a somewhat longer timescale of 9$\times$10$^4$~s.
A constant count rate is ruled out at $>$ 99.9 \% confidence level by the
$\chi^2$ test for the SIS, GIS and MECS data.
On the contrary, no statistically significant deviations  ($<$2$\sigma$) have
been detected in the LECS (0.1-2 keV).
These results suggest that in the 2-10 keV band we are observing the 
intrinsic nuclear X-ray emission, while the
softer component arises from a much larger volume
as, for instance, a scattering medium or a circumnuclear starburst.
In order to check for spectral variations associated with 
the 2-10 keV short term variability we have analyzed the time 
behavior of the hardness ratio 
(hereafter HR) in this range of energy.
We define HR as a function of the 4-10 keV counts (H)
and the 2-4 keV counts (S) as HR=H-S/H+S.
In the upper panel of Fig.~3 and Fig.~4
we show the HR for the GIS and MECS data, respectively.
\begin{figure}
\epsfig{file=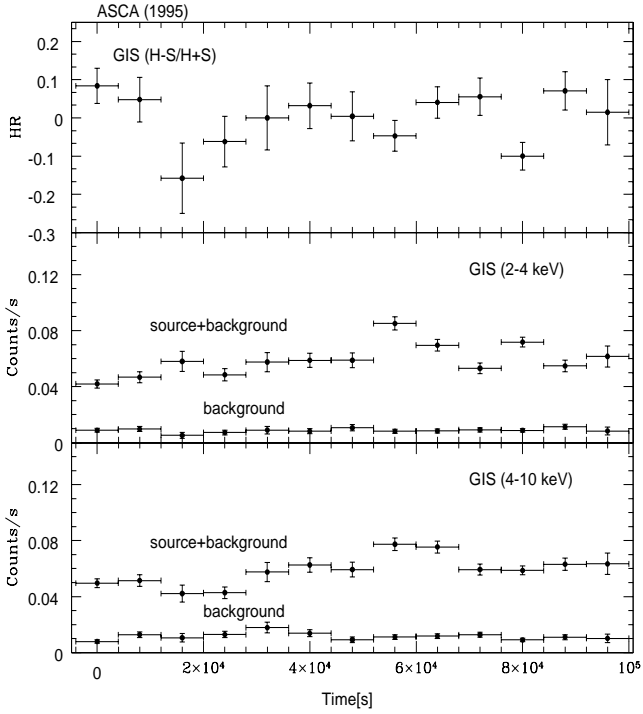, width=9.cm, height=10cm, angle=0}
\caption{Hardness ratio between the 2-4 keV and 4-10 keV bands for the 
GIS data (upper panel). The light curves in each of the two band are shown
in the middle and lower panel. The binning time is 8~ks and the light curves
are not background subtracted.}
\label{ascahr}
\end{figure}
\begin{figure}
\epsfig{file=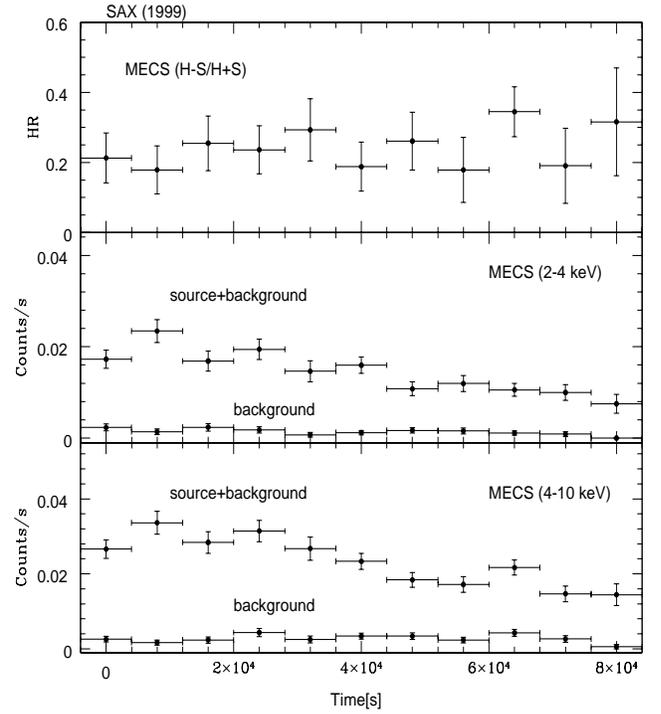, width=9.cm, height=10cm, angle=0}
\caption{Hardness ratio between the 2-4 keV and 4-10 keV bands for the 
MECS data (upper panel). The middle and lower panels show the light curves in 
each of the two energy bands.
The binning time is 8~ks and the light curves
are not background subtracted.}
\label{saxhr}
\end{figure}
For the GIS data a constant HR is marginally rejected by the
data ($>$97\% confidence level according to the $\chi^2$ test),
the MECS data do not show statistically significant variations.

\subsection{Spectral analysis}
The spectral analysis was performed with version 11.0 of XSPEC. The 
spectra were  extracted within an aperture radius of 4 arcmin and
the data were binned in order to achieve 
 a signal-to-noise higher than 3 per channel.
For the ASCA data the background has been extracted in a region within the
same field of the target, paying particularly attention to avoid serendipitous
sources. In particular for the  GIS data we have chosen a region at the same 
off-axis angle of the target.
The ARF files have been created with the ASCAARF ftool, while the RMF files
have been retrieved from the Web page.
On the contrary, for the SAX data we have  used background and calibration 
files provided by the
Science Data Center.
Table 1 shows the exposure time and background subtracted count
rates for the ASCA and Beppo-SAX observations. The errors on the count 
rates are given at the 1$\sigma$ level.
Afterwards, unless otherwise stated, errors will be given at the 90\%
confidence level for one interesting parameter ($\Delta\chi^2$=2.71).

\subsection{ASCA 1995}
The 0.5-10 keV SIS data and the 0.7-10 keV GIS data were simultaneously
fitted in order to obtain the normalization factors between GIS2 and the
other three instruments: 1.21, 0.59 and 0.58 for GIS3, SIS0 and SIS1,
respectively.
The best fit ($\chi^2/dof$=434/400) to a multi-component model typical 
of Compton thin sources is shown in Fig.~5.
\begin{figure}
\epsfig{file=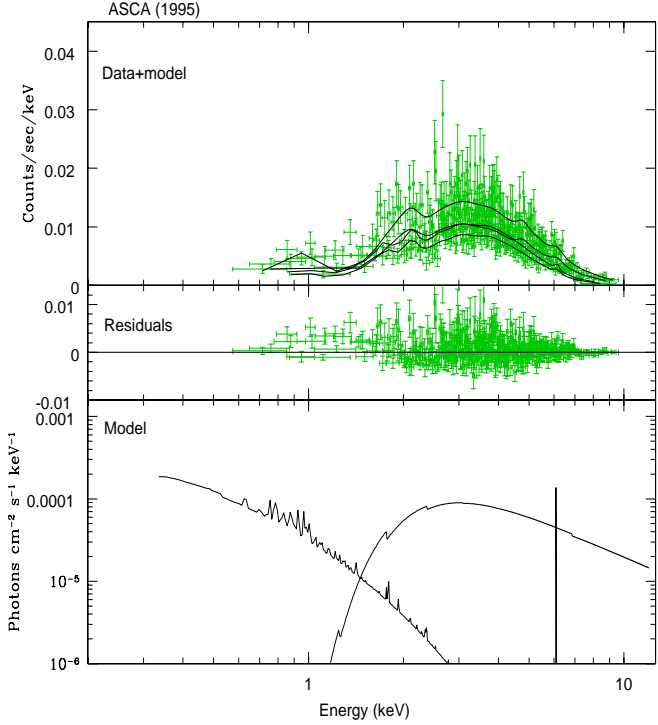, width=9.cm, height=10cm, angle=0}
\caption{Data and folded model (top), residuals (middle) and unfolded model 
(bottom) for the ASCA observation.}
\label{asca_spectrum}
\end{figure}
The relevant best fit parameters  are summarized in the upper
part of Table 2 together with the hard X-ray flux and luminosity derived
by the best fitting model (spectral parameters are quoted in the rest frame).
The soft component is best described by a thermal Raymond-Smith model
with a temperature of 0.87$^{+0.25}_{-0.08}$ keV, while the hard
component (above 2 keV) is compatible with an absorbed powerlaw 
plus Gaussian line model.
The value of the powerlaw photon index (1.71$\pm$0.11) is typical 
of Seyfert galaxies (1.7--1.9; Nandra \& Pounds \cite{nan:pou}, Nandra et al. 
\cite{nan}) and the photoelectric cutoff, corresponding to a  column
density of cold absorbing material of N$_H$=4.36$\times$10$^{22}$ cm$^{-2}$, 
is typical of Compton thin Seyfert~2 galaxies (Bassani et al. \cite{bas}).
An unresolved iron line fixed at E=6.4 keV with an equivalent width 
of EW=113$^{+70}_{-77}$ eV improves the fit at more than 95\%  
confidence level according to the F-test ($\Delta$$\chi^2$/$\Delta$$dof$=6/1).
The line is unresolved, but the upper limit to its width
($\sigma_{Fe}$$<$0.9 keV at 90\% confidence level for one interesting 
parameter) argues in favor of it being originated in a distant torus
rather than a relativistic disk.\\
The SIS to GIS normalization factors which we find are larger than those
commonly expected for the two instruments but are still in the tail of the 
measured distribution (ASCA Helpdesk).
In order to verify that the SIS to GIS mismatch does not affect
our results we have fitted the GIS and SIS spectra separately.
The two sets of best fit parameters are both in full agreement with
those listed in Table~2.

\subsection{Beppo-SAX 1999}

\begin{figure}
\epsfig{file=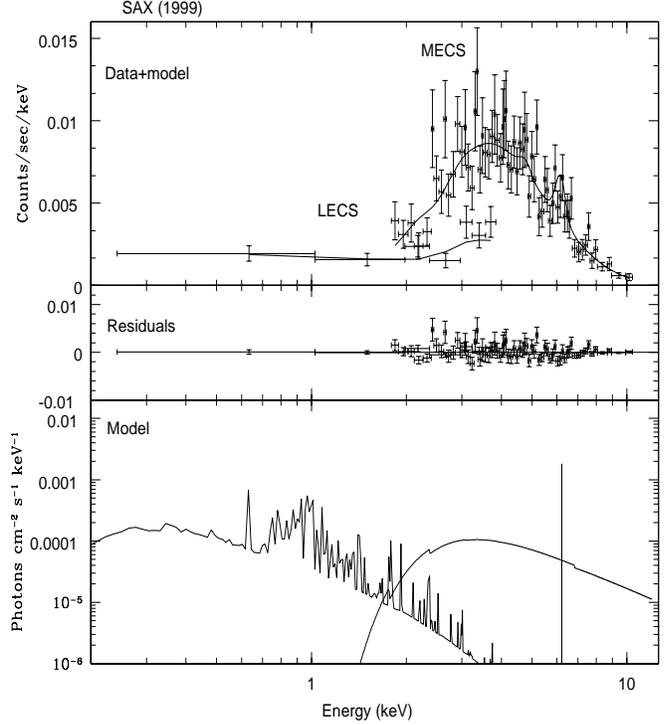, width=9.cm, height=10cm, angle=0}
\caption{Data and folded model (top), residuals (middle) and unfolded model 
(bottom) for the Beppo-SAX observation.}
\label{sax_spectrum}
\end{figure}
\begin{figure}
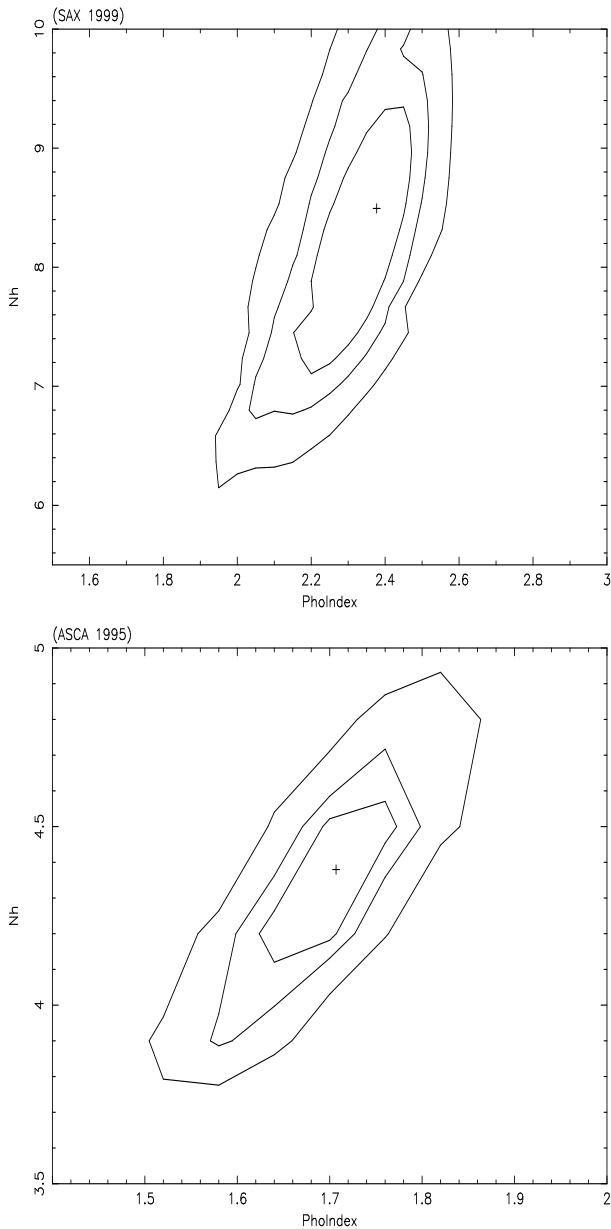

\epsfig{file=fig7a.ps, width=8.cm, height=8cm, angle=-90}
\vskip 0.2 truecm
\epsfig{file=fig7b.ps, width=8.cm, height=8cm, angle=-90}
\caption{Confidence contours of the photon index versus the column density
(in units of 10$^{22}$ cm$^{-2}$) for the Beppo-SAX (upper panel) and
 ASCA (lower panel) observations.
The contour levels are 68\%, 90\% and 99\% for two interesting parameters.}
\label{contours}
\end{figure}
The spectral data from LECS and MECS  were fitted simultaneously
in the 0.1-3.5 keV and 1.8-10.5 keV bands, respectively.
The simultaneous fit of LECS and  MECS in the overlapping band 1.8-3.5 keV
provides a model-independent LECS to MECS normalization factor of 0.77.
The spectrum has been fitted with the same 
three-component model used for the ASCA data,
a Raymond-Smith plasma plus an absorbed powerlaw and an unresolved iron line
fixed at 6.4 keV.
The fit ($\chi^2/dof$=76/70) is shown in Fig.~6 and the best fit parameters 
are listed in Table~2.
The parameters describing the Fe~line and the low energy thermal component
are compatible within the errors with the ones derived from the ASCA data,
again indicating large emission volumes for these components.
Also in this case, the addition of a line component  
improves the
fit at more than 95\% confidence level ($\Delta$$\chi^2$/$\Delta$$dof$=5/1).\\
On the contrary, the continuum above 
2 keV shows strong variability between the Beppo-SAX and ASCA epochs.
The Beppo-SAX data are well reproduced by a steeper powerlaw 
($\Gamma$=2.38, higher than the typical value for Seyfert galaxies)
 and by a photoelectric cutoff corresponding
to a larger column density (N$_H$=8.5$\times$10$^{22}$ cm$^{-2}$).\\
We have verified the significance of the long term spectral shape variation
by calculating the confidence contours for the column density versus the
photon index for both datasets (Fig.~7).
The spectral parameters of 1995 are different from the ones of 1999 at 
a confidence level higher than 99.9\%. 
We have also investigated the possibility that the spectral
variation could arise from a variation of the thermal component.
We have fitted our spectra between 2 and 10 keV only,
and we have found again the same best fit parameters listed in Table~2.
\begin{table*}
\caption{Spectral fits to the ASCA and Beppo-SAX data}
\label{resultt}
\begin{tabular}{ccccccc}
\hline \hline
~~~~~~~~~\\
Model&   kT&$\Gamma$& $N_{\rm H}$& $EW_{\rm K\alpha}$&$f_{\rm 2-10\,keV}^a$
& $L_{\rm 2-10\,keV}^b$\\
&  [keV]& &$[10^{22}cm^{-2}]$ &[eV] &$[10^{-11}erg~s^{-1}cm^{-2}]$& $[10^{42} erg~s^{-1}]$\\ 
~~~~~~~~~\\
\hline
\hline
~~~~~~~~~\\
ASCA(1995)& 0.87$^{+0.25}_{-0.08}$&1.71$\pm0.11$& 4.36$\pm0.42$ & 
113$^{+70}_{-77}$ & 0.48$\pm$0.2 & 19.2$\pm$4.8\\
~~~~~~~~~\\
\hline
\hline
~~~~~~~~~\\
Beppo-SAX(1999)& 1.06$^{+0.34}_{-0.11}$& 2.38$^{+0.09}_{-0.12}$
& 8.5$^{+0.85}_{-0.76}$ & 140$^{+200}_{-99}$&0.36$\pm$0.04& 20.7$\pm$1.28 \\
~~~~~~~~~\\
\hline
\hline
\end{tabular}
~~~~~~~~~\\
~~~~~~~~~\\
$^a$ Luminosities are corrected for the absorption.\\
$^b$ Observed fluxes\\
The errors on fluxes and luminosities are given at the 1$\sigma$ level.
\end{table*}


\section {Near-Infrared spectroscopic  data}
\subsection{Observations and data reduction}
Near Infrared spectroscopic observations of IRAS 05189-2524 
were performed at the ESO New Technology Telescope (NTT).
The data were collected on 1999 November 26 
using the two low resolution grisms available on SOFI:
the Blue Grism (GBF) and the Red Grism (GRF). The Blue and the Red grisms
yield respectively a dispersion of  7 \AA/pix in the 0.95--1.64 $\mu$m
range and 10 \AA/pix in the 1.52--2.52 $\mu$m range. The pixel size is 
of 0.29 arcsec/pix along the 1\arcsec ~ slit and the resolving power was
$\sim$ 500 at 1.25 $\mu$m and $\sim$ 600 at 2.2 $\mu$m.  
The target was observed six times in two different positions along the slit.
This allows us to remove most of the night sky emission by subtracting 
frames from one another.
The exposure time was 50 s on-chip, resulting in a total 
integration time of 300 s on-source with each grism.\\
After sky subtraction each frame was flat fielded with a spectroscopic 
dome exposure.
Wavelength calibration and correction for optical distortion along the 
slit direction were performed using a Xenon arc exposure.
Residual sky emission was removed by fitting a polynomial along
the slit; the spectra thus obtained were combined through a median filter
to have a higher signal to noise and to remove bad pixels.
In order to remove the telluric absorption features, which dominate
the IR spectra, we have used the Infrared spectroscopic standard 
Hip25190 (G5V).
In order to remove from the standard spectrum the intrinsic stellar features 
and to flatten the continuum, we have divided it for a 
synthetic spectrum modeled with the same features and the same slope
of the standard (Sun spectrum).
We have then aligned and divided the IRAS spectrum for the
spectrum resulting from the above process.\\
Since the FWHM of the point spread function along the slit is about 
0.6$\arcsec$, the stellar flux lost in the 1$\arcsec$ slit aperture is 
less than 1\%, and we have used the same standard star to perform
the flux calibration of the spectrum.
The final spectrum was extracted in a 1$\arcsec$$\times$2$\arcsec$
aperture centered on the nucleus of IRAS 05189-2524, and is shown in 
Fig.~8.

\subsection {Lines analysis and reddening estimate}
The NIR hydrogen lines (specifically Pa$\alpha$ and Pa$\beta$)
are characterized by a prominent broad component as clearly shown
by the comparison between their profiles and the profile of the
forbidden [SIII] line in Fig.~9.
For each detected and identified emission line (Fig.~8) a
low order continuum has been fitted to points on both
sides of the line, and then subtracted. 
The uncertainties in the continuum level underneath the lines are included
in the estimate of the errors.
The lines have been fitted with one or two Gaussians.
The model line fluxes were summed and compared with the measured total
line flux and the small discrepancies (less than 5\%) were included 
in the errors.
The values of the central wavelength,
equivalent and physical width, and flux were obtained for each of the lines.
The measurements are listed in Table~3 with typical
uncertainties on EW, FWHM and flux of  6-7\% at the 1$\sigma$ level.
\begin{table}[ht]
\caption{Optical and near--infrared emission lines detected in our spectra.
The typical uncertainties on equivalent widths, line widths and fluxes 
are of 6-7\% at the 1$\sigma$ level.} 
\label{opt_ir_lines}
\begin{tabular}{lcccc}
\hline \hline
Identification & $\lambda_{\rm obs}^a$ &  EW$^b$  & FWHM$^c$ & Flux$^d$ \\
\hline
$[$SIII$]$$\lambda$9531        &  9921 & 15    & 1000   & 4 \\
unident.                       & 10043 &  1    &  600   & 0.2 \\
$[$CI$]$$\lambda$9850          & 10266 &  5    & 1300   & 1.1 \\
HeI$\lambda$10028              & 10451 &  4    & 1100   & 0.9 \\
HeII$\lambda$10123             & 10547 &  4    & 1700   & 1.2 \\
$[$NI$]$$\lambda$10397+10398   & 10833 &  2    & 700    & 0.6 \\
unident.                       & 10933 &  2    & 800    & 0.7 \\
HeI$\lambda$10830+Pa$\gamma$$\lambda$10938     & 11283  & 49  & 4500   & 14 \\
OI$\lambda$11290               & 11767 &  7    & 2600   & 2 \\
$[$FeII$]$$\lambda$12567       & 13098 &  3    & 800    & 1 \\
Pa$\beta$~$\lambda$12818$^{bl}$ & 13354 & 11  & 2100   & 3.8 \\
Pa$\beta$~$\lambda$12818$^{nl}$ & 13359 & 5  & 950    & 1.7 \\
$[$FeII$]$$\lambda$16435       & 17133 & 4     & 800    & 2 \\
Br11$\lambda$16806             & 17508 & 2     & 600    & 0.9 \\
Pa$\alpha$~$\lambda$18751$^{bl}$  & 19528 & 22   & 1900 & 10 \\
Pa$\alpha$~$\lambda$18751$^{nl}$  & 19548 & 15   & 930  & 7 \\
HeI$\lambda$19393               & 20239 & 10   & 2300   & 4 \\
H2(1-0)S(3)~$\lambda$19576      & 20396 & 10   & 1300   & 4 \\
H2(1-0)S(2)~$\lambda$20332      & 21188 & 4    & 1300   & 2 \\
H2(1-0)S(1)~$\lambda$21218      & 22118 & 4    & 1000   & 2 \\
Br$\gamma$~$\lambda$21655       & 22565 & 6    & 1500   & 2\\
HeII$\lambda$22159              & 23067 & 4    & 1300   & 2 \\
H2(1-0)S(0)~$\lambda$22227      & 23170 & 3    & 1100   & 1 \\
\hline \hline
\end{tabular}
\\
$^a$ Observed wavelength in \AA.\\
$^b$ Observed equivalent width in \AA.\\
$^c$ Observed line width in km/s.\\ 
$^d$ Integrated flux in units of $\rm 10^{-14}erg~cm^{-2}~s^{-1}$.\\
$^{bl}$ Broad line component. \\
$^{nl}$ Narrow line component. \\
\end{table}
The amount of reddening which affects the BLR can be calculated by
comparing the observed and intrinsic ratio of the NIR 
broad hydrogen recombination lines.
Indeed, although for the optical hydrogen lines 
radiative transport and collisional effects in the BLR clouds
could affect the standard line ratio expected for case B recombination, 
the intrinsic ratio between NIR lines are much more stable, therefore
 providing a reliable estimate of E$_{B-V}$ (see for instance Drake \& 
Ulrich \cite{dr:ul}).
We will use Pa$\alpha$ and Pa$\beta$ since they have
the highest S/N ratio in our spectra.
Adopting an intrinsic Pa$\alpha$/Pa$\beta$ ratio of 2.057
(case B recombination) and using 
the standard Galactic extinction curve of Savage \& Mathis (\cite{sav}),
we obtain E$_{B-V}$=0.7 mag with a 1$\sigma$ error of $\pm$0.3 mag.
In order to estimate the E$_{B-V}$ value at epochs preceding 1999
we have used the results of Veilleux et al. \cite{vel1}.
These authors present a NIR spectrum of IRAS~05189-2524 obtained in 1997
at UKIRT and they list
the flux intensities of the broad components of both 
Pa$\alpha$ (F=19.5$\times$10$^{-14}$ cm$^{-2}$ s$^{-1}$, 
FWHM=2711 km s$^{-1}$, EW=44 \AA)
and Pa$\beta$ (F=7.34$\times$10$^{-14}$ cm$^{-2}$ s$^{-1}$, 
FWHM=1687 km s$^{-1}$, EW=20 \AA). 
In their spectra (apart from a very broad component of Pa$\beta$
not detected by us) Pa$\alpha$ and Pa$\beta$ are 
unresolved. Their fluxes and equivalent widths should be compared with
our total values (broad plus narrow components), and indeed there is
agreement between the two sets of measurements.
Following the same assumptions described above, we compute from the
results of Veilleux et al. (\cite{vel1}) E$_{B-V}$=0.7 mag with an 
uncertainty of about $\pm$50\%, that is consistent with our value
within the errors.
\section{Discussion}
\subsection{Spectral shape variation}
As discussed in Sect. 2, a large spectral variation is 
observed between the  ASCA and Beppo-SAX epochs.
Our three-parameter model requires that N$_H$ and $\Gamma$ vary together;
however, one could assume more complex scenarios, for instance a 
multi-component absorber, and the evolution of N$_H$ and $\Gamma$ could be
decoupled. A full discussion of this point is not warranted by the quality
of the data, but it is important to stress that in any case 
matter along the line of sight must have
undergone a significant change over a time scale of a few years, with
a large increase in N$_H$  and no corresponding
change in optical absorption.

\subsection {Starburst activity}
\begin{figure*}
\epsfig{file=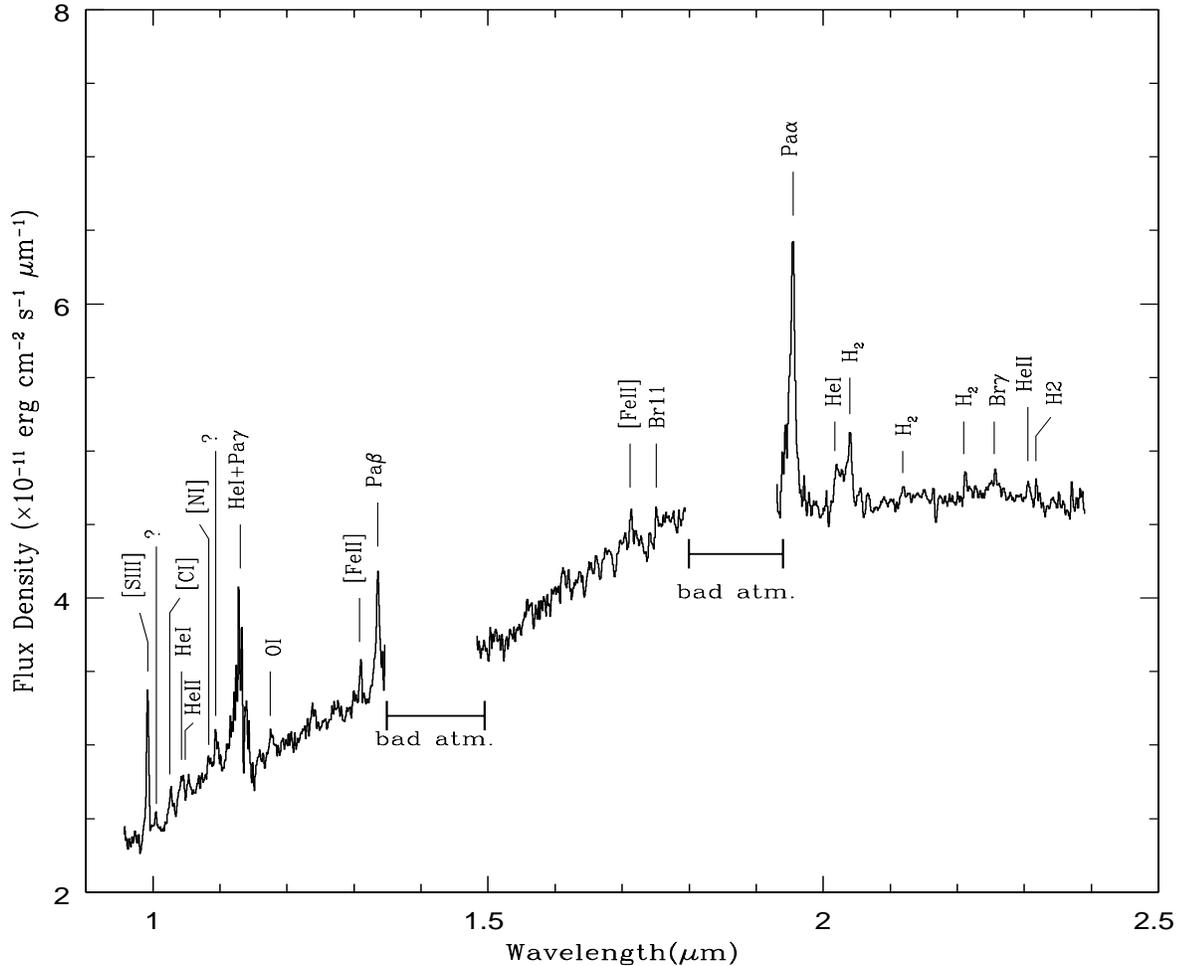, width=17cm, height=14cm, angle=0}
\caption{Near infrared spectrum of IRAS~05189-2524.}
\label{ir_spectrum}
\end{figure*}

Most of the studies carried out so far about IRAS~05189-2524
(e.g. Soifer et al. \cite{soi}, Veilleux et al. \cite{vel1},
Imanishi et al. \cite{ima}) suggest
that the total energy output is predominantly powered by AGN activity.
On the contrary, we find evidence that 
a dominant starburst contribution is strongly required to account for 
the high infrared emission.
Indeed, by taking into account the mean bolometric correction of
Elvis et al. (\cite{elv}, L$_{BOL}$/L$_X$ $\simeq$ 25) and the intrinsic
L$_X$ derived in this paper, we estimate an AGN bolometric luminosity  
L$_{BOL}$$\simeq$1.4$\times$10$^{11}$ L$_\odot$.
This value is about 10 times lower than the total IR luminosity 
of this source (L$_{IR}$$\simeq$1.3$\times$10$^{12}$ L$_\odot$),
thus implying the presence of a dominant starburst component.\\
Our result is also supported by the work of
 Risaliti et al. (\cite{ris}).
They present a study of the hard X--ray properties of 
all the LIGs observed in the 2--10 keV band, and find a 
clear correlation between their F$_X$/F$_{IR}$ ratio and the
infrared color (25/60 $\mu$m).
In particular, moving towards lower 25/60 $\mu$m they find lower
X/IR ratios and an increasing fraction of obscured AGNs at first,
and of starburts afterwards.
Their model reproduces the IR-X correlation 
by means of mixed AGN and starburst contributions.
The differences in IR colors are mainly due to the different
contribution of the starburst component, while the X/IR flux ratio is
mainly determined by the amount of absorption affecting the AGN.
Following this model, the infrared and X-ray properties of our source 
are in agreement with those of a starburst-dominated object,
therefore confirming the starburst dominance inferred above by
the comparison of L$_{BOL}$ and L$_X$.\\
Finally, the presence of a starburst component is  also in agreement
with the thermal component required by the spectral fits.

\subsection {Dust-to-Gas ratio}
As anticipated in Sect.~1, there are various observational evidences
suggesting that the dust reddening (E$_{B-V}$) and the absorption (A$_V$)
towards AGNs and ULIRGs is lower than the values expected from the gaseous
column density (N$_H$) measured in the X rays (assuming a Galactic 
dust-to-gas ratio and extinction curve). 
Using our results for
IRAS~05189-2524, we obtain E$_{B-V}$/N$_H$=8.2$\times$10$^{-24}$ mag cm$^2$ 
and E$_{B-V}$/N$_H$=1.6$\times$10$^{-23}$ mag cm$^2$ assuming N$_H$ 
equal to 8.5$\times$10$^{22}$ cm$^{-2}$ and 4.36$\times$10$^{22}$ cm$^{-2}$,
respectively. Both values are much lower than the Galactic standard value
1.7$\times$10$^{-22}$ mag cm$^2$ (Bohlin et al. \cite{boh}), and this
result remains true when the errors on E$_{B-V}$ and N$_H$ are taken
into account. A low dust-to-gas ratio can be ascribed to various effects, 
as described in Maiolino et al. \cite{mai2}.
Since the nucleus of this source shows a significant polarization (Young et al.
\cite{you}), one possibility is that the broad lines used to measure
E$_{B-V}$ are scattered by a reflecting mirror observed along
a line of sight with a column density much lower than the X--ray source.
In this hypothesis, taking into account that the
scattering efficiency generally does not exceed a few percent,
we would expect that the reddening-corrected line luminosities are 
underluminous with respect to the absorption-corrected X--ray luminosity,
when compared to unobscured Seyfert~1 galaxies and QSOs.
With the redenning value estimated in this paper we have 
corrected the observed flux of the broad Pa$\alpha$ component and we have
derived the intrinsic luminosity of the broad H$\alpha$ line as
L$_{H\alpha}$=5.6$\times$10$^{42}$ erg s$^{-1}$
(we have adopted an intrinsic Pa$\alpha$/H$\alpha$ ratio of 0.106, case B 
recombination, and the Galactic extinction curve of
Savage \& Mathis \cite{sav}). This is the typical $H\alpha$ luminosity 
that we expect to find in an unobscured AGN with an intrinsic
L$_X$ similar to that of our ULIRG (see Fig.~1 of Maiolino et al. 
\cite{mai2}). If we take H$\alpha$ with the estimated intrinsic luminosity,
and attenuate it through the A$_V$ determined from the NIR lines,
we predict an observed flux of 1.7$\pm$1.5$\times$10$^{-13}$
erg s$^{-1}$ cm$^{-2}$, fully consistent with the broad component
measured by Young et al. (\cite{you}) (3.1$\times$10$^{-13}$ erg s$^{-1}$
cm$^{-2}$). Thus, not only the NIR broad lines, but also the H$\alpha$
broad component is seen in transmission. The scattering model
suggested by Young et al. (\cite{you}) is not supported by our results, and 
a type 1.9 classification is more appropriate than the usual type 2.0
given in the literature.\\
Other interpretations must be invoked to explain the low dust-to-gas
ratio measured in this source.
The obscuring material could be characterized by a lower total amount
of dust or, alternatively, the dust grains could be larger than in the
Galactic ISM, producing a flatter extinction curve 
(Laor and Draine \cite{lao}, Maiolino et al. \cite{mai1}, \cite{mai2}).
\begin{figure}
\epsfig{file=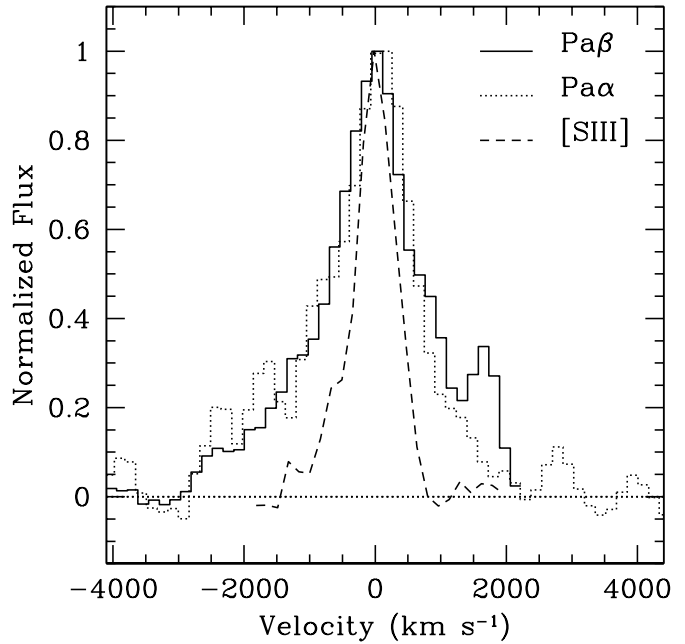, width=9cm, height=9cm, angle=0}
\caption{Comparison between Pa$\alpha$ (dotted line),
Pa$\beta$ (solid line) and [SIII] (dashed line) profiles.}
\label{profiles}
\end{figure}

\section{Conclusion}
In this paper we have discussed X--ray spectra (ASCA 1995 and Beppo-SAX 1999)
and NIR spectra (UKIRT 1997 and SOFI-NTT 1999) of the ULIRG IRAS~05189-2524.
Above 2 keV this is a typical Compton-thin Seyfert galaxy.
Comparing the spectral parameters obtained with ASCA and Beppo-SAX,
 we find a long term spectral shape variation.
The absorbing matter along the line of sight
must have undergone a major change between the two epochs (1995 and 1999),
 with a large increase in N$_H$.
On the contrary, from the broad transmitted lines detected in the NIR
spectrum, no significant change in the optical absorption has been
revealed, thus supporting a correlation between the (non standard) 
properties of the
obscuring matter  and its proximity to the center.
Comparing the A$_V$ and N$_H$ values we find a dust-to-gas ratio
definitely less than the Galactic one, even if N$_H$ is given the lower
value measured in the previous years.
The value of A$_V$ measured in this paper is compatible with 
the broad component of H$\alpha$ being seen in transmission, contrary to
what  found previously by Young et al. (\cite{you}). This implies that 
the high degree of polarization observed in this source must
 be due to dichroic transmission and leads to a more appropriate  1.9 Seyfert 
classification.
Finally, using the results of our analysis coupled with the IR data retrieved 
from the literature, we find that the bolometric luminosity of the AGN can not
account for the high IR emission, thus implying the presence of a 
dominant starburst component.
\begin{acknowledgements}
We acknowledge the financial support of the Italian Ministry of University and
Research (MURST) under grant Cofin-98-02-32.
We thank Roberto Gilli and Cristian Vignali for helpful comments.
\end{acknowledgements}

\end{document}